# *Electromagnetic Scattering by Bianisotropic Spheres*

Maxim Durach*

Department of Physics and Astronomy, Georgia Southern University, Statesboro, Georgia 30461, USA

Corresponding author: *mdurach@georgiasouthern.edu

**Abstract**

Electromagnetic fields in bulk bianisotropic media can be represented using plane waves whose k-vectors can be found using the *index of refraction operator method* and belong to the Fresnel wave surfaces that fall into one of the 5 hyperbolic classes that are used as the *taxonomy of bianisotropic media* [Durach et al., Appl. Sci., Opt. Comm. (2020), PIER (2022)]. It has been demonstrated that, alternatively, the linear combinations of vector spherical harmonics can be used as a set of solutions of vector Helmholtz equation in gyroelectric, gyromagnetic, and gyrotropic anisotropic media to develop Mie theory of scattering by anisotropic spheres [Lin, Chui, Phys. Rev. E (2004), Li, Ong, Zheng, Phys. Rev. E (2012)]. In this paper we introduce electromagnetic orbitals for bianisotropic media as linear combinations of vector spherical harmonics, which represent a set of solutions of Maxwell's equations in bianisotropic media. Using these *bianisotropic orbitals* we develop a theory of scattering of electromagnetic radiation by bianisotropic spheres with arbitrary effective material parameters and sizes. As a by-product we obtain a simple expression for the expansion of a vector plane wave over vector spherical harmonics (cf. Sarkar, Halas, Phys. Rev. E (1997)). We obtain the polarizability expressions in Rayleigh limit of our theory in agreement with the previous results of the electrostatic approximation [Lakhtakia, J. Phys (1990), Sihvola, Mic. Opt. Tech. Lett. (1994)].

## ***1. Introduction***

Electromagnetism occupies the crowned role in physics, science, and modern technology. As in the cases of the 2nd and the 3rd Industrial Revolutions, the research in electromagnetism is driving the ongoing 4th Industrial Revolution [1] related to the transition to renewable energy, telecommunications in 5G and 6G standards [2], advanced micro-/nanofabrication for novel electronic devices [3], bioelectromagnetics [4], information and electronic warfare [5], machine learning, material training [6-7], and in other realms. The slithering scientific and technological unification of the physical, chemical, biological, and digital worlds brought by the 4th Industrial Revolution is due to the inalienable electromagnetic nature of these phenomena, based in the rule of the underlying Coulomb's, Gauss's, Biot-Savart's, Ampere's, Kirchhoff's, Faraday's, and Maxwell's Laws [8-10]. We the researchers of the modern electromagnetism are devoted to the development of the new electromagnetic materials, collectively called composite artificial materials or metamaterials [11-17]. Metamaterials are made of arrays of subwavelength scatterers designed to exhibit the desired electromagnetic properties. In many important cases metamaterials can be described as bianisotropic media [13-17]. In bianisotropic media both the electric and magnetic responses depend on both the electric and magnetic fields of the external radiation [18,19].

The studies of bianisotropic materials are almost as old as the electromagnetism itself persisting through the 19th and 20th centuries in the work of such scientists as Roentgen, Wilson, Landau, Lifshitz, Dzyaloshinskii, Cheng, and Kong [18-24]. In the 21st century the field of bianisotropic optical materials has received the name of bianisotropics [25-27] and is closely related to the electromagnetic metamaterial research, since, typically, the desired properties of metamaterials depend on them being anisotropic and bianisotropic media [13-19]. Despite all these efforts and rich history of research into bianisotropics until recently very few general concrete properties of bianisotropic media were established beyond formulaic apparatus due to the complexity of these materials and multiparametric nature of these media [14,18,28-35]. The bianisotropic media are the most general case of local linear media [18,19,25,36,37] with the effective material parameters combined into a 6x6 material parameters matrix $\widehat{M}$, which characterized the electric displacement field $\boldsymbol{D}$ and magnetic field $\boldsymbol{B}$ in terms of the fields $\boldsymbol{E}$ and $\boldsymbol{H}$:

$$\begin{pmatrix}\boldsymbol{D}\\ \boldsymbol{B}\end{pmatrix} = \widehat{M}\begin{pmatrix}\boldsymbol{E}\\ \boldsymbol{H}\end{pmatrix} = \begin{pmatrix}\hat{\epsilon} & \widehat{X}\\ \widehat{Y} & \hat{\mu}\end{pmatrix}\begin{pmatrix}\boldsymbol{E}\\ \boldsymbol{H}\end{pmatrix}. \quad (1)$$

The 3x3 matrices $\hat{\epsilon}, \hat{\mu}, \widehat{X}, \widehat{Y}$ are dielectric permittivity, magnetic permeability and two magnetoelectric coupling matrices respectively. The inverse relationship can also be formulated:

$$\begin{pmatrix}\boldsymbol{E}\\ \boldsymbol{H}\end{pmatrix} = \begin{bmatrix}\hat{\epsilon} & \widehat{X}\\ \widehat{Y} & \hat{\mu}\end{bmatrix}^{-1}\begin{pmatrix}\boldsymbol{D}\\ \boldsymbol{B}\end{pmatrix} = \begin{bmatrix}\hat{\alpha}_{ED} & \hat{\alpha}_{EB}\\ \hat{\alpha}_{HB} & \hat{\alpha}_{HD}\end{bmatrix}\begin{pmatrix}\boldsymbol{D}\\ \boldsymbol{B}\end{pmatrix}. \quad (2)$$

One of the jewels in the crown of electromagnetism is Mie theory of scattering by spheres. Originally, Mie theory was introduced to describe scattering by isotropic spheres [38,39], but later it was extended to describe scattering by biisotropic [40], rotationally-symmetric anisotropic [41], orthorhombic dielectric–magnetic [42], magneto- and electro-gyrotropic [43-45], and spherically-symmetric bianisotropic [46] spheres. Despite of this activity, the theory of scattering by a generic bianisotropic sphere has not yet been constituted [47].

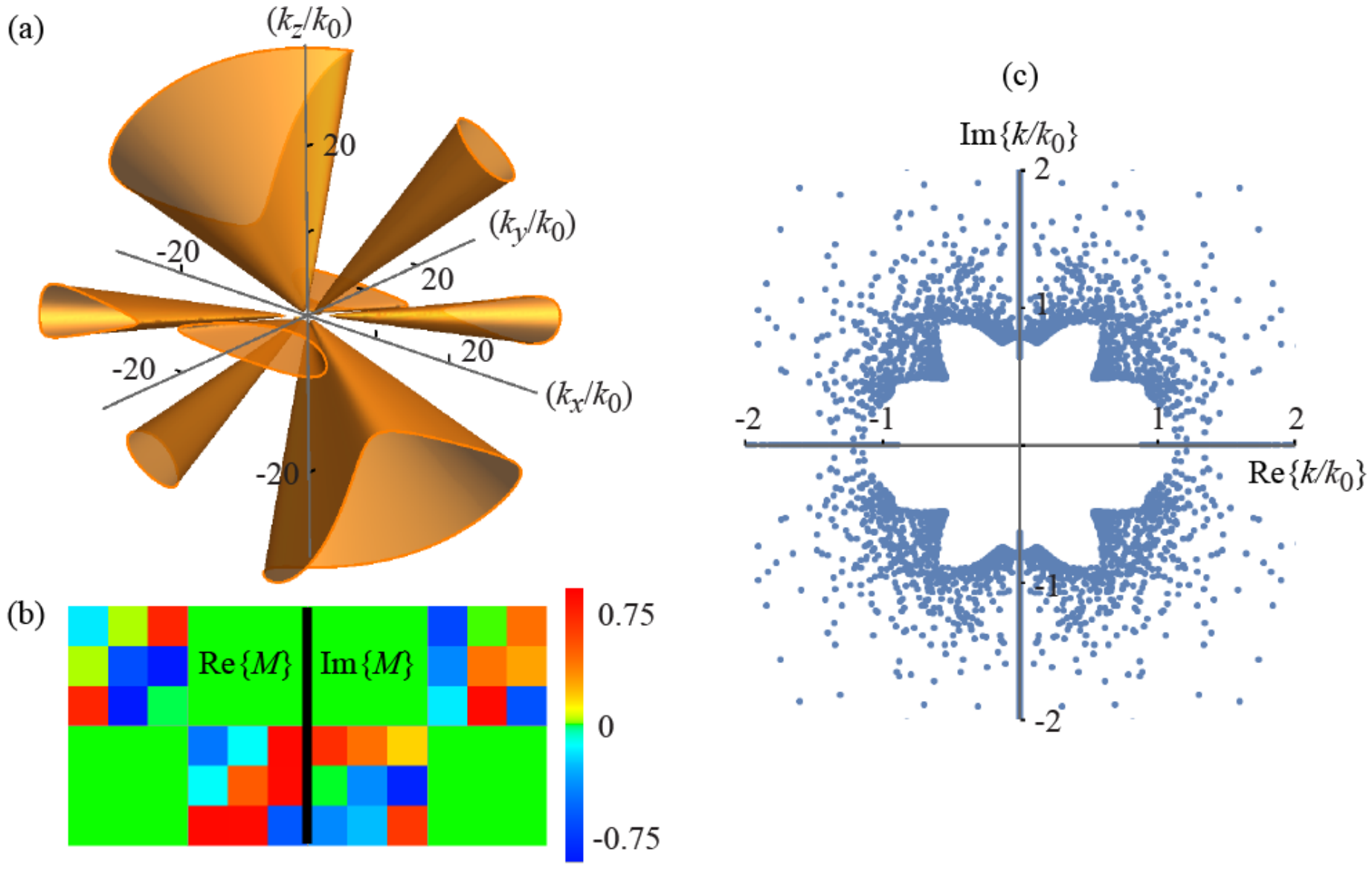

Fig. 1. Iso-frequency Fresnel wave surface for a tetra-hyperbolic bianisotropic medium with the effective material parameters matrix $\widehat{M}$ color-coded in panel (b); (c) eigenvalues of the index of refraction operator [31-33] in complex plane for the same material.

The plane waves which can propagate in bianisotropic media belong to Fresnel wave surfaces, which can be characterized using the index of refraction operator method [31-33]. The Fresnel wave surfaces in bianisotropic media follow quartic dispersion equations, and, therefore, can be classified using the Durach et al. taxonomy [31-33] which includes the 5 hyperbolic classes: non-, mono-, bi-, tri-, and tetra-hyperbolic materials [31-33, 48-59]. The prefix in the name of each topological class indicates the number of the double cones that the iso-frequency k-surface has in the high-k limit. Note that hyperbolic metamaterials, already known for applications in optical imaging, hyperlensing, and emission rate and directivity control utilizing the diverging optical density of high-k states [51-59]. In Fig. 1(a) we show an example of an iso-frequency Fresnel wave surface for a tetra-hyperbolic bianisotropic medium with the effective material parameters matrix $\widehat{M}$ color-coded in Fig. 1(b).

The Fresnel wave surfaces only include the plane waves with real wave vectors $\boldsymbol{k}$. Nevertheless, the inhomogeneous plane waves with imaginary $\boldsymbol{k}$ are also important. In Fig. 1(c) we plot the eigenvalues of the index of refraction operator [31-33] in complex plane for the material color-coded in Fig. 1(b) for plane waves propagating in directions spanning the entire solid angle with steps in angles $\Delta\theta = \pi/40$ and $\Delta\phi = 2\pi/40$.

In unbounded homogeneous bianisotropic media plane waves represent the set of solutions of Maxwell's equations which is typically used, but to consider scattering of electromagnetic waves by bianisotropic spheres it is more convenient to express the electromagnetic fields inside of such spheres in terms of vector spherical harmonics. In this paper we propose a theory of scattering by bianisotropic spheres with arbitrary effective media parameters $\widehat{M}$. To accomplish this, we introduce bianisotropic orbitals composed of vector spherical harmonics.

## 2. *Bianisotropic orbitals*

The plane waves whose k-vectors belong to Fresnel's wave surfaces represent a complete set of solutions of Maxwell's equations in the bulk bianisotropic media. A different set of solutions can be found as expansion over vector spherical harmonics. Recently, the close connection between the multipole composition of electromagnetic fields and bianisotropy has been revealed [60].

Due to the solenoidal nature of $\boldsymbol{D}$- and $\boldsymbol{B}$-fields: $\nabla \cdot \boldsymbol{D} = 0$ and $\nabla \cdot \boldsymbol{B} = 0$, we can express them as an expansion over the vector spherical harmonics $\boldsymbol{M}_{lm}^{1}, \boldsymbol{N}_{lm}^{1}$ found from solutions of the scalar Helmholtz equations [61-63]:

$$\psi_{lm}^{(j)} = -\frac{1}{i\sqrt{l(l+1)}} z_l^{(j)}(kr) Y_{lm},\ \psi_{00}^{(j)} = i z_0^{(j)}(kr) Y_{00} \tag{3}$$

according to

$$\boldsymbol{L}_{lm}^{(j)} = \frac{1}{k}\nabla\psi_{lm}^{(j)},\ \boldsymbol{M}_{lm}^{(j)} = \nabla \times \left(\boldsymbol{r}\psi_{lm}^{(j)}\right),\ \boldsymbol{N}_{lm}^{(j)} = \frac{1}{k}\nabla \times \boldsymbol{M}_{lm}^{(j)} \tag{4}$$

(the detailed definitions of the vector spherical harmonics used in this paper is in Appendix A).

We represent the $\boldsymbol{D}$- and $\boldsymbol{B}$-fields of a *bianisotropic orbital* with a wave number $k_q$ as

$$\boldsymbol{D}_q = \sum_{lm}\left\{f_{qlm}^{DM}\boldsymbol{M}_{lm}^{1}(k_q) + f_{qlm}^{DN}\boldsymbol{N}_{lm}^{1}(k_q)\right\} \tag{5}$$

$$\boldsymbol{B}_q = \sum_{lm}\left\{f_{qlm}^{BM}\boldsymbol{M}_{lm}^{\boldsymbol{1}}(k_q) + f_{qlm}^{BN}\boldsymbol{N}_{lm}^{\boldsymbol{1}}(k_q)\right\} \tag{6}$$

The corresponding $\boldsymbol{E}$- and $\boldsymbol{H}$-fields can be expressed using Eq. (2) as

$$\boldsymbol{E}_q = \sum_{uv}\left(\mu_{uv}^{eq}\boldsymbol{M}_{uv}^{1q} + \nu_{uv}^{eq}\boldsymbol{N}_{uv}^{1q} + \lambda_{uv}^{eq}\boldsymbol{L}_{uv}^{1q}\right) = \hat{\alpha}_{ED}\boldsymbol{D}_q + \hat{\alpha}_{EB}\boldsymbol{B}_q$$
$$= \sum_{lm}\left\{f_{qlm}^{DM}\left(\hat{\alpha}_{ED}\boldsymbol{M}_{lm}^{1q}\right) + f_{qlm}^{DN}\left(\hat{\alpha}_{ED}\boldsymbol{N}_{lm}^{1q}\right)\right\} + \sum_{lm}\left\{f_{qlm}^{BM}\left(\hat{\alpha}_{EB}\boldsymbol{M}_{lm}^{1q}\right) + f_{qlm}^{BN}\left(\hat{\alpha}_{EB}\boldsymbol{N}_{lm}^{1q}\right)\right\}$$
$$= \sum\nolimits_{lm,uv}\left(\boldsymbol{M}_{uv}^{1q}, \boldsymbol{N}_{uv}^{1q}, \boldsymbol{L}_{uv}^{1q}\right) \cdot \begin{pmatrix} g_{MM}^{\alpha_{ED}} & g_{NM}^{\alpha_{ED}} & g_{MM}^{\alpha_{EB}} & g_{NM}^{\alpha_{EB}} \\ g_{MN}^{\alpha_{ED}} & g_{NN}^{\alpha_{ED}} & g_{MN}^{\alpha_{EB}} & g_{NN}^{\alpha_{EB}} \\ g_{ML}^{\alpha_{ED}} & g_{NL}^{\alpha_{ED}} & g_{ML}^{\alpha_{EB}} & g_{NL}^{\alpha_{EB}} \end{pmatrix}_{lm,uv} \cdot \begin{pmatrix} f_{qlm}^{DM} \\ f_{qlm}^{DN} \\ f_{qlm}^{BM} \\ f_{qlm}^{BN} \end{pmatrix} \tag{7}$$

$$\boldsymbol{H}_q = \sum_{uv}\left(\mu_{uv}^{hq}\boldsymbol{M}_{uv}^{1q} + \nu_{uv}^{hq}\boldsymbol{N}_{uv}^{1q} + \lambda_{uv}^{hq}\boldsymbol{L}_{uv}^{1q}\right) = \hat{\alpha}_{HD}\boldsymbol{D}_q + \hat{\alpha}_{HB}\boldsymbol{B}_q$$
$$= \sum_{lm}\left\{f_{qlm}^{DM}\left(\hat{\alpha}_{HD}\boldsymbol{M}_{lm}^{1q}\right) + f_{qlm}^{DN}\left(\hat{\alpha}_{HD}\boldsymbol{N}_{lm}^{1q}\right)\right\} + \sum_{lm}\left\{f_{qlm}^{BM}\left(\hat{\alpha}_{HB}\boldsymbol{M}_{lm}^{1q}\right) + f_{qlm}^{BN}\left(\hat{\alpha}_{HB}\boldsymbol{N}_{lm}^{1q}\right)\right\}$$
$$= \sum\nolimits_{lm,uv}\left(\boldsymbol{M}_{uv}^{1q}, \boldsymbol{N}_{uv}^{1q}, \boldsymbol{L}_{uv}^{1q}\right) \cdot \begin{pmatrix} g_{MM}^{\alpha_{HD}} & g_{NM}^{\alpha_{HD}} & g_{MM}^{\alpha_{HB}} & g_{NM}^{\alpha_{HB}} \\ g_{MN}^{\alpha_{HD}} & g_{NN}^{\alpha_{HD}} & g_{MN}^{\alpha_{HB}} & g_{NN}^{\alpha_{HB}} \\ g_{ML}^{\alpha_{HD}} & g_{NL}^{\alpha_{HD}} & g_{ML}^{\alpha_{HB}} & g_{NL}^{\alpha_{HB}} \end{pmatrix}_{lm,uv} \cdot \begin{pmatrix} f_{qlm}^{DM} \\ f_{qlm}^{DN} \\ f_{qlm}^{BM} \\ f_{qlm}^{BN} \end{pmatrix} \tag{8}$$

where coefficients $g$ are found using $\{\boldsymbol{U}|\hat{\alpha}|\boldsymbol{V}\} = \int_0^\infty \int_0^{2\pi} \int_0^\pi \boldsymbol{U} \cdot \hat{\alpha} \cdot \boldsymbol{V}\, r^2 \sin\theta\, dr\, d\theta\, d\varphi$ , $\int Y_{uv}^*(\hat{\boldsymbol{r}})Y_{lm}(\hat{\boldsymbol{r}})d\Omega = \delta_{ul}\delta_{vm}$, and $\int_0^\infty j_u(k'r)j_l(kr)r^2dr = \frac{\pi}{2k^2}\delta(k-k')$ as:

$$\begin{pmatrix} \{\boldsymbol{M}_{uv}^\star|\hat{\alpha}|\boldsymbol{M}_{lm}\} & \{\boldsymbol{N}_{uv}^\star|\hat{\alpha}|\boldsymbol{M}_{lm}\} & \{\boldsymbol{L}_{uv}^\star|\hat{\alpha}|\boldsymbol{M}_{lm}\} \\ \{\boldsymbol{M}_{uv}^\star|\hat{\alpha}|\boldsymbol{N}_{lm}\} & \{\boldsymbol{N}_{uv}^\star|\hat{\alpha}|\boldsymbol{N}_{lm}\} & \{\boldsymbol{L}_{uv}^\star|\hat{\alpha}|\boldsymbol{N}_{lm}\} \\ \{\boldsymbol{M}_{uv}^\star|\hat{\alpha}|\boldsymbol{L}_{lm}\} & \{\boldsymbol{N}_{uv}^\star|\hat{\alpha}|\boldsymbol{L}_{lm}\} & \{\boldsymbol{L}_{uv}^\star|\hat{\alpha}|\boldsymbol{L}_{lm}\} \end{pmatrix} =$$
$$= \begin{pmatrix} g_{MM}^{\alpha} & g_{MN}^{\alpha} & g_{ML}^{\alpha} \\ g_{NM}^{\alpha} & g_{NN}^{\alpha} & g_{NL}^{\alpha} \\ g_{LM}^{\alpha} & g_{LN}^{\alpha} & g_{LL}^{\alpha} \end{pmatrix}_{lm,uv} \begin{pmatrix} \{\boldsymbol{M}_{uv}^\star|\boldsymbol{M}_{uv}\} = 1 & 0 & 0 \\ 0 & \{\boldsymbol{N}_{uv}^\star|\boldsymbol{N}_{uv}\} = 1 & 0 \\ 0 & 0 & \{\boldsymbol{L}_{uv}^\star|\boldsymbol{L}_{uv}\} \end{pmatrix} \tag{9}$$

The *bianisotropic orbitals* with radial quantum numbers $k_q$ represent solutions of Maxwell's equations in homogeneous bianisotropic media if the expansion coefficients $f_{qlm}$ satisfy the following eigenproblem:

$$\sum\nolimits_{lm} \begin{pmatrix} -g_{MN}^{\alpha_{HD}} & -g_{NN}^{\alpha_{HD}} & -g_{MN}^{\alpha_{HB}} & -g_{NN}^{\alpha_{HB}} \\ -g_{MM}^{\alpha_{HD}} & -g_{NM}^{\alpha_{HD}} & -g_{MM}^{\alpha_{HB}} & -g_{NM}^{\alpha_{HB}} \\ g_{MN}^{\alpha_{ED}} & g_{NN}^{\alpha_{ED}} & g_{MN}^{\alpha_{EB}} & g_{NN}^{\alpha_{EB}} \\ g_{MM}^{\alpha_{ED}} & g_{NM}^{\alpha_{ED}} & g_{MM}^{\alpha_{EB}} & g_{NM}^{\alpha_{EB}} \end{pmatrix}_{lm,uv} \begin{pmatrix} f_{qlm}^{DM} \\ f_{qlm}^{DN} \\ f_{qlm}^{BM} \\ f_{qlm}^{BN} \end{pmatrix} = i\left(\frac{k_0}{k_q}\right) \begin{pmatrix} f_{quv}^{DM} \\ f_{quv}^{DN} \\ f_{quv}^{BM} \\ f_{quv}^{BN} \end{pmatrix} \tag{10}$$

Note that the eigenproblem of Eq. (10) differs from the eigenproblems formulated for the anisotropic media, which stem from the vector Helmholtz equations [43-45]. Such eigenproblems can not be used in the case of bianisotropic media considered here and we use the pair of Maxwell's equations composed of Faraday's and Maxwell-Ampere's equations instead to obtain Eq. (10).

The *bianisotropic orbitals* provided by solutions of Eq. (10) can be represented as expansions over plane wave solutions of the index of refraction operator method [31-33] with the indexes of refraction $n = k_q/k_0$ (Appendix B). In Fig. 2(a) we show the inverse eigenvalues $k_q/k_0$ of Eq. (10) for the eigenproblems which are truncated and $l = 4$ (black dots), 10 (red), and 40 (green). Note the direct correspondence of the eigenvalues of Eq. (10) in Fig. 2(a) with the eigenvalues of the index of refraction operator plotted in Fig. 1(b). This correspondence between the plane waves and the bianisotropic orbitals introduced in this paper shows the relationship between the solution of the scattering problem presented in this paper with the method of plane wave expansion proposed for scattering by the uniaxial anisotropic spheres [64,65]. In Fig. 2(b-e) we plot the components $|f_{qlm}|^2$ of the eigenproblem of Eq. (10) for a bianisotropic orbital with $k_q/k_0 = 2.2$ in the angular momentum $l$-$m$ space.

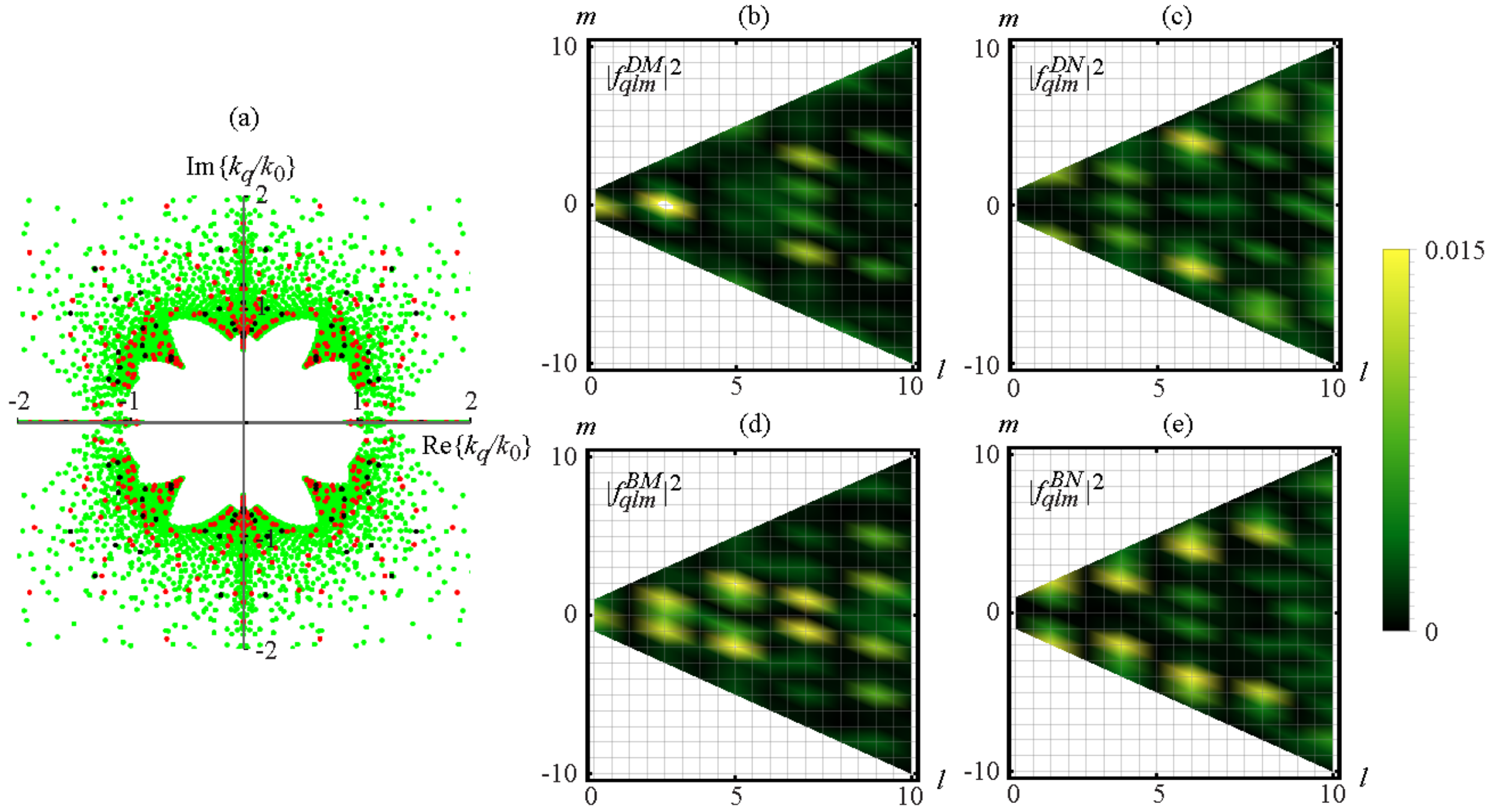

Fig. 2 (a) inverse eigenvalues $k_q/k_0$ of the eigenproblem in Eq. (10) truncated at $l = 4$ (black dots), 10 (red dots), 40 (green dots). Note the correspondence with the eigenvalues of the index of refraction operator plotted in Fig. 1(b); (b)-(e) eigenvector components $|f_{qlm}|^2$ of the eigenproblem Eq. (10) for a bianisotropic orbital with $k_q/k_0 = 2.2$ in the angular momentum $l$-$m$ space.

### *3. Scattering cross-section of bianisotropic spheres in vacuum*

The bianisotropic orbitals can be used to solve a large range of problems with spherically shaped bianisotropic media from spheres to spherical shells, to spherical voids, to combinations of such geometries and so forth. Here we consider a bianisotropic sphere with arbitrary effective material parameters $\widehat{M}$ with radius $R$. We study scattering of an electromagnetic plane wave by such sphere. The field of the plane wave is given by

$$\boldsymbol{E}_{in} = \boldsymbol{\epsilon}(\boldsymbol{k})e^{i\boldsymbol{kr}} = \sum_{lm}\left(q_{lm}\boldsymbol{M}^1_{lm} + p_{lm}\boldsymbol{N}^1_{lm}\right) \tag{11}$$

$$\boldsymbol{H}_{in} = \boldsymbol{h}(\boldsymbol{k})e^{i\boldsymbol{kr}} = \frac{1}{i}\sum_{lm}\left(p_{lm}\boldsymbol{M}^1_{lm} + q_{lm}\boldsymbol{N}^1_{lm}\right), \ \boldsymbol{h} = \widehat{\boldsymbol{k}} \times \boldsymbol{\epsilon} \tag{12}$$

We define the orientation of the incident electric and magnetic fields with polarization angle $\alpha$ in the spherical coordinates with respect to the incidence direction $\hat{\boldsymbol{k}} = (\sin\theta\cos\phi, \sin\theta\sin\phi, \cos\theta)$, $\boldsymbol{\epsilon}(\boldsymbol{k}) = (\sin\alpha\,\hat{\boldsymbol{\theta}} - \cos\alpha\,\hat{\boldsymbol{\phi}})$, $\boldsymbol{h}(\boldsymbol{k}) = (\cos\alpha\,\hat{\boldsymbol{\theta}} + \sin\alpha\,\hat{\boldsymbol{\phi}})$, where $\hat{\boldsymbol{\theta}} = (\cos\theta\cos\phi, \cos\theta\sin\phi, -\sin\theta)$, $\hat{\boldsymbol{\phi}} = (-\sin\phi, \cos\phi)$.

Vector plane wave expansions over vector spherical harmonics exist in literature [43,66]. Nevertheless, we derive a compact expansion of a vector plane wave for our work (Appendix C):

$$\boldsymbol{a}e^{i\boldsymbol{kr}} = 4\pi\sum_{lm} i^l\left(-\sqrt{l(l+1)}\left\{\boldsymbol{a}\cdot\boldsymbol{Y}_{lm}^{(-1)*}(\hat{\boldsymbol{k}})\right\}\boldsymbol{L}_{lm}^{(1)} + \{\boldsymbol{a}\cdot\boldsymbol{Y}_{lm}^{l*}(\hat{\boldsymbol{k}})\}\boldsymbol{M}_{lm}^{(1)} - \left\{\boldsymbol{a}\cdot\boldsymbol{Y}_{lm}^{(+1)*}(\hat{\boldsymbol{k}})\right\}\boldsymbol{N}_{lm}^{(1)}\right)$$

Correspondingly, the coefficients in the expansions Eqs. (11)-(12) are

$$q_{lm} = 4\pi i^l\{\boldsymbol{\epsilon}(\boldsymbol{k})\cdot\boldsymbol{Y}_{lm}^{l*}(\hat{\boldsymbol{k}})\},\ p_{lm} = 4\pi i^{l+1}\{\boldsymbol{h}(\boldsymbol{k})\cdot\boldsymbol{Y}_{lm}^{l*}(\hat{\boldsymbol{k}})\} \tag{13}$$

The scattered fields outside of the sphere are given by

$$\boldsymbol{E}_{sc} = \sum_{lm}\left(b_{lm}\boldsymbol{M}_{lm}^3 + a_{lm}\boldsymbol{N}_{lm}^3\right) \tag{14}$$

$$\boldsymbol{H}_{sc} = \frac{1}{i}\sum_{lm}\left(a_{lm}\boldsymbol{M}_{lm}^3 + b_{lm}\boldsymbol{N}_{lm}^3\right) \tag{15}$$

We represent the field inside of the bianisotropic sphere as a linear combination of *bianisotropic orbitals* described by Eq. (10).

$$\boldsymbol{E}_{sph} = \sum_q A_q\boldsymbol{E}_q, \quad \boldsymbol{H}_{sph} = \sum_q A_q\boldsymbol{H}_q, \tag{16}$$

Please note that for reciprocal media the eigenvalues of Eq. (10) come in pairs with radial wavenumber $\pm k_q$, and the corresponding modes in Eq. (16) are equivalent. Therefore, among the orbitals with real $k_q$ we only include the modes with positive $k_q$. For the orbitals with complex $k_q$ we only include the modes with $\mathrm{Im}\,k_q > 0$. Note, however, that in non-reciprocal media the reciprocity symmetry between modes is broken, and all the bianisotropic orbitals of Eq. (10) should be included into Eq. (16). Correspondingly, additional boundary conditions (ABC) are needed as described in Ref. [34].

The continuity of tangential components of E- and H-fields at the surface of the sphere lead to the following boundary conditions expressed in terms of Riccati-Bessel functions $\psi_l(x) = xj_l(x)$ and $\xi_l(x) = xh_l^{(1)}(x)$, where $j_l(x)$ and $h_l^{(1)}(x)$ are spherical Bessel function of the first and third kinds, and parameters $x = k_0R$ and $x_q = k_qR$:

$$E_{in\theta} + E_{sc\theta} = E_{sph\theta},\quad q_{lm} + b_{lm}\left(\frac{\xi_l(x)}{\psi_l(x)}\right) = \sum_q A_q\left(\frac{x}{x_q}\right)\mu_{lm}^{eq}\left(\frac{\psi_l(x_q)}{\psi_l(x)}\right) \tag{17}$$

$$H_{in\theta} + H_{sc\theta} = H_{sph\theta},\quad p_{lm} + a_{lm}\left(\frac{\xi_l(x)}{\psi_l(x)}\right) = i\sum_q A_q\left(\frac{x}{x_q}\right)\mu_{lm}^{hq}\left(\frac{\psi_l(x_q)}{\psi_l(x)}\right) \tag{18}$$

$$E_{in\phi} + E_{sc\phi} = E_{sph\phi},\quad p_{lm} + a_{lm}\left(\frac{\xi_l'(x)}{\psi_l'(x)}\right) = \sum_q A_q\left(\frac{x}{x_q}\right)\left\{\nu_{lm}^{eq}\left(\frac{\psi_l'(x_q)}{\psi_l'(x)}\right) + \lambda_{lm}^{eq}\left(\frac{j_l(x_q)}{\psi_l'(x)}\right)\right\} \tag{19}$$

$$H_{in\phi} + H_{sc\phi} = H_{sph\phi},\quad q_{lm} + b_{lm}\left(\frac{\xi_l'(x)}{\psi_l'(x)}\right) = i\sum_q A_q\left(\frac{x}{x_q}\right)\left\{\nu_{lm}^{hq}\left(\frac{\psi_l'(x_q)}{\psi_l'(x)}\right) + \lambda_{lm}^{hq}\left(\frac{j_l(x_q)}{\psi_l'(x)}\right)\right\} \tag{20}$$

The boundary equations (17)-(20) can be represented in the matrix form as

$$\hat{M}_\psi\vec{A} - \hat{\Psi}(ab) = (pq) \tag{21}$$

$$\widehat{N}_\psi \vec{A} - \widehat{\Phi}(ab) = (pq) \tag{22}$$

where $\vec{A}$ is the column of the amplitudes of the *bianisotropic orbitals* $A_q$ inside the sphere, $(ab)$ and $(pq)$ include the amplitudes of the scattered vector spherical harmonics and the known incident amplitudes of Eq. (13).

Matrices $\widehat{\Psi}$, $\widehat{\Phi}$, are diagonal and contain the functions $\left(\frac{\xi_l(x)}{\psi_l(x)}\right)$ and $\left(\frac{\xi_l'(x)}{\psi_l'(x)}\right)$, while matrices $\widehat{M}_\psi$, $\widehat{N}_\psi$ represent the coupling between different multipoles due to the bianisotropy in the bianisotropic orbitals and correspond to the sums in the right-hand sides of Eqs. (17)-(20)

Excluding $(pq)$ coefficients we get the relationship between $(ab)$ and $\vec{A}$

$$(ab) = \widehat{\Omega}\vec{A},\ \ \widehat{\Omega} = \left\{\widehat{\Psi} - \widehat{\Phi}\right\}^{-1}\left\{\widehat{M}_\psi - \widehat{N}_\psi\right\} \tag{23}$$

The amplitudes of the *bianisotropic photonic orbitals* $A_q$ are found from Eqs. (21) and (23) as

$$\vec{A} = \widehat{\Xi}(pq) = \left\{\widehat{M}_\psi - \widehat{\Psi}\,\widehat{\Omega}\right\}^{-1}(pq) \tag{24}$$

Substituting the amplitudes $A_q$ from Eq. (24) into Eq. (23) we obtain the T-matrix for a bianisotropic sphere with arbitrary effective medium parameters and the scattering amplitudes $(ab)$ in terms of the parameters of the incident wave $(pq)$ as

$$(ab) = \widehat{T}(pq),\ \ \widehat{T} = \widehat{\Omega}\widehat{\Xi} \tag{25}$$

The scattering cross-section $Q_s$ can be found from the scattering amplitudes as

$$Q_s = \frac{1}{k_0^2}\sum_{lm}\left(|a_{lm}|^2 + |b_{lm}|^2\right) \tag{26}$$

In Fig. 3(a-b) we plot the dependence of the reduced scattering cross-section $Q_s/(\pi R^2)$ for all incidence angles $\theta$ and $\phi$ for a bianisotropic sphere with the effective material parameters color-coded in Fig. 1(b) and $x = k_0 R = 1$. Panel Fig. 3(a) corresponds to the incidence polarization angle $\alpha = 0$, while in panel (b) the polarization angle is $\alpha = \pi/2$.

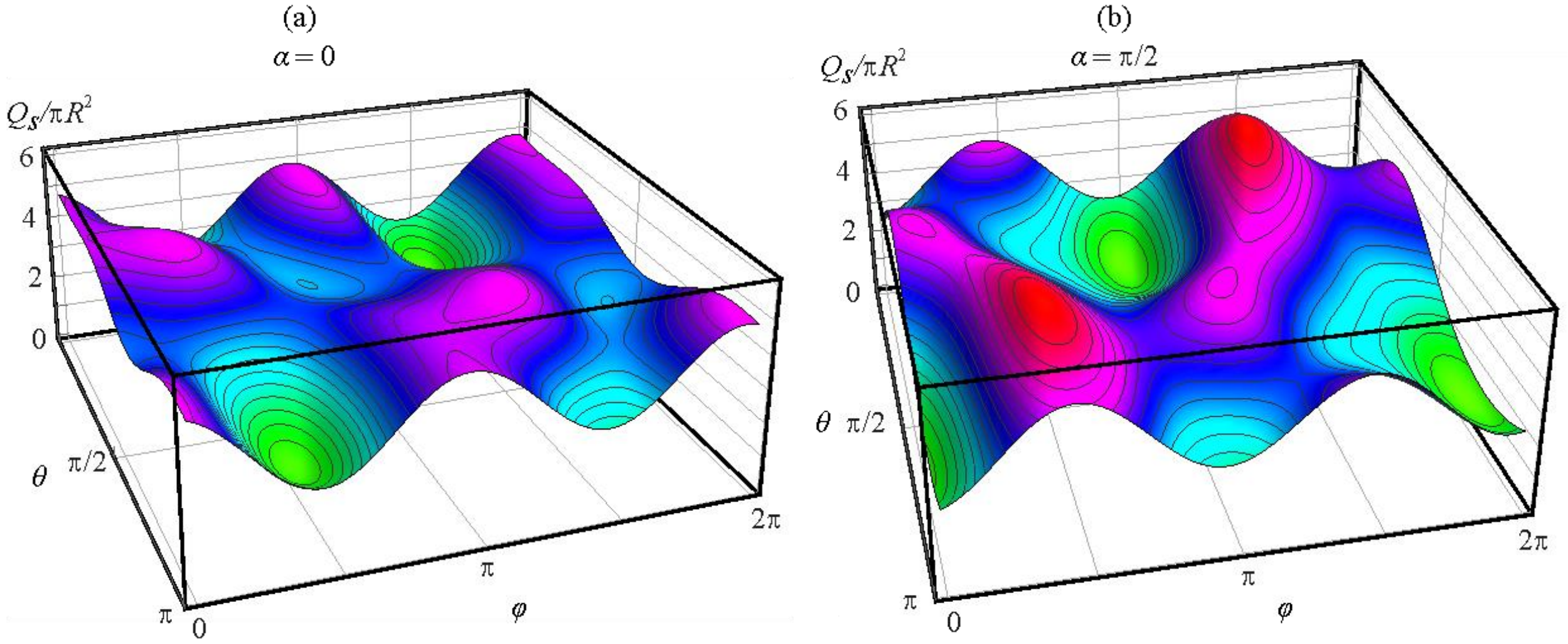

Fig. 3. Reduced scattering cross-section $Q_s/(\pi R^2)$ for all incidence angles $\theta$ and $\phi$ for a bianisotropic sphere with the effective material parameters color-coded in Fig. 1(b) and $x = k_0 R = 1$; In panel (a) incidence polarization angle $\alpha = 0$; in (b) $\alpha = \pi/2$

In Fig. 3 one can see a strong dependence of scattering by bianisotropic spheres on incidence direction and polarization.

### *4. Polarizability expressions in the Rayleigh limit*

In Rayleigh limit of electromagnetically small spheres $x \ll 1$ only the dipole terms of the Mie theory are retained. For $l, u = 1$, $g_{MN}^{\alpha} = g_{ML}^{\alpha} = g_{NM}^{\alpha} = 0$ and the $\boldsymbol{E}$- and $\boldsymbol{H}$-fields Eqs. (7)-(8) can be written as

$$\boldsymbol{E} = \sum_{mv} \boldsymbol{M}_{1v}^{1}\left(g_{MM}^{\alpha ED}, g_{MM}^{\alpha EB}\right)_{m,v} \begin{pmatrix} f_{1m}^{DM} \\ f_{1m}^{BM} \end{pmatrix} + \sum_{mv} \boldsymbol{N}_{1v}^{1}\left(g_{NN}^{\alpha ED}, g_{NN}^{\alpha EB}\right)_{m,v} \begin{pmatrix} f_{1m}^{DN} \\ f_{1m}^{BN} \end{pmatrix} + \sum_{mv} \boldsymbol{L}_{1v}^{1}\left(g_{NL}^{\alpha ED}, g_{NL}^{\alpha EB}\right)_{m,v} \begin{pmatrix} f_{1m}^{DN} \\ f_{1m}^{BN} \end{pmatrix} \quad (27)$$

$$\boldsymbol{H} = \sum_{mv} \boldsymbol{M}_{1v}^{1}\left(g_{MM}^{\alpha HD}, g_{MM}^{\alpha HB}\right)_{m,v} \begin{pmatrix} f_{1m}^{DM} \\ f_{1m}^{BM} \end{pmatrix} + \sum_{mv} \boldsymbol{N}_{1v}^{1}\left(g_{NN}^{\alpha HD}, g_{NN}^{\alpha HB}\right)_{m,v} \begin{pmatrix} f_{1m}^{DN} \\ f_{1m}^{BN} \end{pmatrix} + \sum_{mv} \boldsymbol{L}_{1v}^{1}\left(g_{NL}^{\alpha HD}, g_{NL}^{\alpha HB}\right)_{m,v} \begin{pmatrix} f_{1m}^{DN} \\ f_{1m}^{BN} \end{pmatrix} \quad (28)$$

Applying Maxwell equation to $\boldsymbol{E}$- and $\boldsymbol{H}$-fields of Eqs. (20), (21) we get

$$-i(k_0/k)\boldsymbol{D} = \sum_{1m,1v} \boldsymbol{N}_{1v}^{1} \cdot \left(g_{MM}^{\alpha HD}, g_{MM}^{\alpha HB}\right)_{m,v} \cdot \begin{pmatrix} f_{1m}^{DM} \\ f_{1m}^{BM} \end{pmatrix} + \sum_{1m,1v} \boldsymbol{M}_{1v}^{1} \cdot \left(g_{NN}^{\alpha HD}, g_{NN}^{\alpha HB}\right)_{m,v} \cdot \begin{pmatrix} f_{1m}^{DN} \\ f_{1m}^{BN} \end{pmatrix}$$

$$i(k_0/k)\boldsymbol{B} = \sum_{1m,1v} \boldsymbol{N}_{1v}^{1} \cdot \left(g_{MM}^{\alpha ED}, g_{MM}^{\alpha EB}\right)_{m,v} \cdot \begin{pmatrix} f_{1m}^{DM} \\ f_{1m}^{BM} \end{pmatrix} + \sum_{1m,1v} \boldsymbol{M}_{1v}^{1} \cdot \left(g_{NN}^{\alpha ED}, g_{NN}^{\alpha EB}\right)_{m,v} \cdot \begin{pmatrix} f_{1m}^{DN} \\ f_{1m}^{BN} \end{pmatrix}$$

which translates into a system of equations

$$\hat{G}_M F_{Mq} = i(k_0/k_q)\hat{U} F_{Nq} \quad (29)$$

$$\hat{G}_N F_{Nq} = i(k_0/k_q)\hat{V} F_{Mq} \quad (30)$$

where $F_{Mq} = \left(f_{1-1q}^{DM}, f_{1-1q}^{BM}, f_{10q}^{DM}, f_{10q}^{BM}, f_{11q}^{DM}, f_{11q}^{BM}\right)^T$, $F_{Mq} = \left(f_{1-1q}^{DN}, f_{1-1q}^{BN}, f_{10q}^{DN}, f_{10q}^{BN}, f_{11q}^{DN}, f_{11q}^{BN}\right)^T$, and matrices $\hat{G}_M$ and $\hat{G}_N$ are composed of coefficients $g_{MM}$ and $g_{NN}$

$$\hat{G}_M = \begin{pmatrix} g_{MM--}^{\alpha HD} & g_{MM--}^{\alpha HB} & g_{MM0-}^{\alpha HD} & g_{MM0-}^{\alpha HB} & g_{MM+-}^{\alpha HD} & g_{MM+-}^{\alpha HB} \\ g_{MM--}^{\alpha ED} & g_{MM--}^{\alpha EB} & g_{MM0-}^{\alpha ED} & g_{MM0-}^{\alpha EB} & g_{MM+-}^{\alpha ED} & g_{MM+-}^{\alpha EB} \\ g_{MM-0}^{\alpha HD} & g_{MM-0}^{\alpha HB} & g_{MM00}^{\alpha HD} & g_{MM00}^{\alpha HB} & g_{MM+0}^{\alpha HD} & g_{MM+0}^{\alpha HB} \\ g_{MM-0}^{\alpha ED} & g_{MM-0}^{\alpha EB} & g_{MM00}^{\alpha ED} & g_{MM00}^{\alpha EB} & g_{MM+0}^{\alpha ED} & g_{MM+0}^{\alpha EB} \\ g_{MM-+}^{\alpha HD} & g_{MM-+}^{\alpha HB} & g_{MM0+}^{\alpha HD} & g_{MM0+}^{\alpha HB} & g_{MM++}^{\alpha HD} & g_{MM++}^{\alpha HB} \\ g_{MM-+}^{\alpha ED} & g_{MM-+}^{\alpha EB} & g_{MM0+}^{\alpha ED} & g_{MM0+}^{\alpha EB} & g_{MM++}^{\alpha ED} & g_{MM++}^{\alpha EB} \end{pmatrix},$$

$$\hat{G}_N = \begin{pmatrix} g_{NN--}^{\alpha ED} & g_{NN--}^{\alpha EB} & g_{NN0-}^{\alpha ED} & g_{NN0-}^{\alpha EB} & g_{NN+-}^{\alpha ED} & g_{NN+-}^{\alpha EB} \\ g_{NN--}^{\alpha HD} & g_{NN--}^{\alpha HB} & g_{NN0-}^{\alpha HD} & g_{NN0-}^{\alpha HB} & g_{NN+-}^{\alpha HD} & g_{NN+-}^{\alpha HB} \\ g_{NN-0}^{\alpha ED} & g_{NN-0}^{\alpha EB} & g_{NN00}^{\alpha ED} & g_{NN00}^{\alpha EB} & g_{NN+0}^{\alpha ED} & g_{NN+0}^{\alpha EB} \\ g_{NN-0}^{\alpha HD} & g_{NN-0}^{\alpha HB} & g_{NN00}^{\alpha HD} & g_{NN00}^{\alpha HB} & g_{NN+0}^{\alpha HD} & g_{NN+0}^{\alpha HB} \\ g_{NN-+}^{\alpha ED} & g_{NN-+}^{\alpha EB} & g_{NN0+}^{\alpha ED} & g_{NN0+}^{\alpha EB} & g_{NN++}^{\alpha ED} & g_{NN++}^{\alpha EB} \\ g_{NN-+}^{\alpha HD} & g_{NN-+}^{\alpha HB} & g_{NN0+}^{\alpha HD} & g_{NN0+}^{\alpha HB} & g_{NN++}^{\alpha HD} & g_{NN++}^{\alpha HB} \end{pmatrix}$$

while $\hat{U} = \begin{pmatrix} -1 & 0 & 0 & 0 & 0 & 0 \\ 0 & 1 & 0 & 0 & 0 & 0 \\ 0 & 0 & -1 & 0 & 0 & 0 \\ 0 & 0 & 0 & 1 & 0 & 0 \\ 0 & 0 & 0 & 0 & -1 & 0 \\ 0 & 0 & 0 & 0 & 0 & 1 \end{pmatrix}$, $\hat{V} = \begin{pmatrix} 0 & 1 & 0 & 0 & 0 & 0 \\ -1 & 0 & 0 & 0 & 0 & 0 \\ 0 & 0 & 0 & 1 & 0 & 0 \\ 0 & 0 & -1 & 0 & 0 & 0 \\ 0 & 0 & 0 & 0 & 0 & 1 \\ 0 & 0 & 0 & 0 & -1 & 0 \end{pmatrix}$.

Excluding $F_{Mq} = i(k_q/k_0)\hat{V}\hat{G}_N F_{Nq}$ from the system Eqs. (29)-(30) we obtain an eigenproblem for $F_{Nq}$, which corresponds to 6 *bianisotropic orbitals* in Rayleigh limit

$$\left(\hat{U}\hat{G}_M\hat{V}\hat{G}_N\right)F_{Nq} = \left(k_0/k_q\right)^2 F_{Nq} \tag{31}$$

The boundary conditions Eq. (17)-(20) turn into

$$j_1(k_0R)\,(pq) + h_1(k_0R)(ab) = \textstyle\sum_q A_q\, j_1\!\left(k_qR\right)\mathrm{diag}\{i,1,i,1,i,1\}\hat{G}_M F_{Mq} \tag{32}$$

$$\frac{\partial}{\partial r}\left[r\, j_1(k_0r)\right]_{r=R}(pq) + \frac{\partial}{\partial r}\left[r\, h_1(k_0r)\right]_{r=R}(ab) = \sum_q A_q\left(\frac{k_0}{k_q}\right)\mathrm{diag}\{1,i,1,i,1,i\}\times$$

$$\times\left(\frac{\partial}{\partial r}\left[r\, j_1\!\left(k_qr\right)\right]_{r=R}\hat{G}_N F_{Nq} + j_1\!\left(k_qR\right)\hat{G}_L F_{Nq}\right) \tag{33}$$

where $\hat{G}_L$ is composed of coefficients $g_{NL}$ similarly to the matrix $\hat{G}_N$ provided above, $(pq) = (p_{1-1}, q_{1-1}, p_{10}, q_{10}, p_{11}, q_{11})^T$, and $(ab) = (a_{1-1}, b_{1-1}, a_{10}, b_{10}, a_{11}, b_{11})^T$

Taking the limit of $k_qR, k_0R \to 0$ in the spherical Bessel functions, and excluding the scattered amplitudes $(ab)$, and substituting Eq. (29) for $\hat{G}_M F_{Mq}$, we obtain

$$3(pq) = \mathrm{diag}\{1,i,1,i,1,i\}\left(\hat{1} + 2\hat{G}_N + \hat{G}_L\right)\hat{F}_N\vec{A} \tag{34}$$

The Eq. (34) can be re-written using $\begin{pmatrix}\boldsymbol{D}\\ \boldsymbol{B}\end{pmatrix} = \frac{1}{\sqrt{3\pi}}\hat{T}^{-1}\hat{F}_N\vec{A}$, $(pq) = \sqrt{3\pi}\hat{T}\begin{pmatrix}\boldsymbol{\epsilon}\\ \boldsymbol{h}\end{pmatrix}$, the matrix $\hat{T}$ given by $\hat{T} = \begin{pmatrix} -i & +1 & 0 & 0 & 0 & 0\\ 0 & 0 & 0 & 1 & +i & 0\\ 0 & 0 & -i\sqrt{2} & 0 & 0 & 0\\ 0 & 0 & 0 & 0 & 0 & \sqrt{2}\\ i & +1 & 0 & 0 & 0 & 0\\ 0 & 0 & 0 & -1 & +i & 0\end{pmatrix}$

In accordance with Eq. (34), $\left(\hat{1} + 2\hat{G}_N + \hat{G}_L\right) = \hat{T}\left(\hat{1} + 2\hat{M}^{-1}\right)\hat{T}^{-1}$, or $3\hat{T}^{-1}\left(\hat{1} + 2\hat{G}_N + \hat{G}_L\right)^{-1}\hat{T} = 3\hat{M}\left(\hat{M} + 2\hat{1}\right)^{-1}$. This results in

$$\begin{pmatrix}\boldsymbol{D}\\ \boldsymbol{B}\end{pmatrix} = 3\hat{M}\left(\hat{M} + 2\hat{1}\right)^{-1}\begin{pmatrix}\boldsymbol{\epsilon}\\ \boldsymbol{h}\end{pmatrix} \tag{35}$$

Considering, that $\begin{pmatrix}\boldsymbol{D}\\ \boldsymbol{B}\end{pmatrix} = \begin{pmatrix}\boldsymbol{E}\\ \boldsymbol{H}\end{pmatrix} + 4\pi\begin{pmatrix}\boldsymbol{P}\\ \boldsymbol{M}\end{pmatrix} = \hat{M}^{-1}\begin{pmatrix}\boldsymbol{D}\\ \boldsymbol{B}\end{pmatrix} + 4\pi\begin{pmatrix}\boldsymbol{P}\\ \boldsymbol{M}\end{pmatrix}$, we get

$$\begin{pmatrix}\boldsymbol{D}\\ \boldsymbol{B}\end{pmatrix} = 4\pi\left(\hat{1} - \hat{M}^{-1}\right)^{-1}\begin{pmatrix}\boldsymbol{P}\\ \boldsymbol{M}\end{pmatrix},$$

$$\begin{pmatrix}\boldsymbol{P}\\ \boldsymbol{M}\end{pmatrix} = \frac{3}{4\pi}\left(\hat{M} - \hat{1}\right)\left(\hat{M} + 2\hat{1}\right)^{-1}\begin{pmatrix}\boldsymbol{\epsilon}\\ \boldsymbol{h}\end{pmatrix} \tag{36}$$

From this we get the polarizability $\hat{\alpha}$ of the sphere of volume $V$ in Rayleigh approximation

$$\begin{pmatrix}\boldsymbol{p}\\ \boldsymbol{m}\end{pmatrix} = V\begin{pmatrix}\boldsymbol{P}\\ \boldsymbol{M}\end{pmatrix} = \hat{\alpha}\begin{pmatrix}\boldsymbol{\epsilon}\\ \boldsymbol{h}\end{pmatrix} = \frac{3V}{4\pi}\left(\hat{M} - \hat{1}\right)\left(\hat{M} + 2\hat{1}\right)^{-1}\begin{pmatrix}\boldsymbol{\epsilon}\\ \boldsymbol{h}\end{pmatrix}$$

This agrees with the polarizability of the bianisotropic spheres obtained previously in electrostatic approximation [67-69].

***Conclusion***

To conclude, in this paper we presented a theory of scattering by bianisotropic spheres with arbitrary effective media parameters and sizes. In Rayleigh limit we obtained the results known from the electrostatic approximation approach.

**Acknowledgements**

This work was supported by the Georgia Southern University Scholarly Pursuit Funding Award.

**Appendices**

***Appendix A. Vector Spherical Harmonics***

We use a definition of vector spherical harmonics (VSH) that builds upon the definitions of Stratton [61], Jackson [62], and Varshalovich et al. [63]. The starting point is the solution of the scalar Helmholtz equation $\nabla^2\psi + k^2\psi = 0$

$$\psi_{lm}^{(j)} = -\frac{1}{i\sqrt{l(l+1)}} z_l^{(j)}(kr)Y_{lm}, \qquad \psi_{00}^{(j)} = iz_0^{(j)}(kr)Y_{00},$$

where the scalar spherical harmonic is $Y_{lm}(\hat{\boldsymbol{r}}) = \sqrt{\frac{2l+1}{4\pi}\cdot\frac{(l-m)!}{(l+m)!}}P_l^m(\cos\theta)e^{im\varphi}$.

The VSH are derived from $\psi_{lm}^{(j)}$ analogous to the definitions of Stratton [61] as

$$\boldsymbol{L}_{lm}^{(j)} = \frac{1}{k}\nabla\psi_{lm}^{(j)}, \qquad \boldsymbol{M}_{lm}^{(j)} = \nabla\times\left(\boldsymbol{r}\psi_{lm}^{(j)}\right), \qquad \boldsymbol{N}_{lm}^{(j)} = \frac{1}{k}\nabla\times\boldsymbol{M}_{lm}^{(j)}$$

$$\boldsymbol{L}_{00}^{(j)} = \frac{1}{k}\nabla\psi_{00}^{(j)}, \qquad \boldsymbol{M}_{00}^{(j)} = 0, \qquad \boldsymbol{N}_{00}^{(j)} = 0$$

In spherical coordinates these VSH can be expressed as

$$\boldsymbol{L}_{lm}^{(j)} = \frac{1}{k}\nabla\psi_{lm}^{(j)} = \frac{1}{k}\frac{\partial\psi_{lm}^{(j)}}{\partial r}\hat{\boldsymbol{r}} + \frac{1}{kr}\frac{\partial\psi_{lm}^{(j)}}{\partial\theta}\hat{\boldsymbol{\theta}} + \frac{1}{kr}\frac{im}{\sin\theta}\psi_{lm}^{(j)}\hat{\boldsymbol{\varphi}}$$

$$\boldsymbol{M}_{lm}^{(j)} = \nabla\times\left(\boldsymbol{r}\psi_{lm}^{(j)}\right) = \frac{im}{\sin\theta}\psi_{lm}^{(j)}\hat{\boldsymbol{\theta}} - \frac{\partial\psi_{lm}^{(j)}}{\partial\theta}\hat{\boldsymbol{\varphi}}$$

$$\boldsymbol{N}_{lm}^{(j)} = \frac{1}{k}\nabla\times\boldsymbol{M}_{lm}^{(j)} = \frac{l(l+1)}{kr}\psi_{lm}^{(j)}\hat{\boldsymbol{r}} + \frac{1}{kr}\frac{\partial^2}{\partial r\partial\theta}\left[r\psi_{lm}^{(j)}\right]\hat{\boldsymbol{\theta}} + \frac{1}{kr}\frac{im}{\sin\theta}\frac{\partial}{\partial r}[r\psi_{lm}^{(j)}]\hat{\boldsymbol{\varphi}}$$

These harmonics are directly related to the harmonics of Jackson [62]

$$\boldsymbol{X}_{lm}(\hat{\boldsymbol{r}}) = \frac{1}{\sqrt{l(l+1)}}\boldsymbol{L}Y_{lm}, \qquad \hat{\boldsymbol{r}}Y_{lm}, \qquad \hat{\boldsymbol{r}}\times\boldsymbol{X}_{lm},$$

where $\boldsymbol{L} = -i(\boldsymbol{r}\times\nabla)$ is the angular momentum operator as

$$\boldsymbol{M}_{lm}^{(j)} = \frac{z_l^{(j)}(kr)}{i\sqrt{l(l+1)}}\boldsymbol{r}\times\nabla Y_{lm} = z_l^{(j)}(kr)\boldsymbol{X}_{lm}(\hat{\boldsymbol{r}})$$

$$\boldsymbol{N}_{lm}^{(j)} = \frac{1}{k}\nabla \times \left(z_l^{(j)}(kr)\boldsymbol{X}_{lm}(\hat{\boldsymbol{r}})\right) = \frac{i\sqrt{l(l+1)}}{kr}\, z_l^{(j)}(kr)\,\hat{\boldsymbol{r}}Y_{lm} + \frac{1}{kr}\frac{\partial}{\partial r}\left(rz_l^{(j)}(kr)\right)\,\hat{\boldsymbol{r}}\times\boldsymbol{X}_{lm}$$

The vector spherical harmonics $\boldsymbol{L}_{lm}^{(j)}, \boldsymbol{N}_{lm}^{(j)}, \boldsymbol{M}_{lm}^{(j)}$ can be expressed using the harmonics $\boldsymbol{Y}_{jm}^{l}(\hat{\boldsymbol{r}})$ of Varshalovich et al. [63]

$$\begin{pmatrix} \boldsymbol{L}_{lm}^{(j)} \\ \boldsymbol{M}_{lm}^{(j)} \\ \boldsymbol{N}_{lm}^{(j)} \end{pmatrix} = \begin{pmatrix} \frac{iz_{l-1}^{(j)}(kr)}{\sqrt{(l+1)(2l+1)}} & 0 & \frac{iz_{l+1}^{(j)}}{\sqrt{l(2l+1)}} \\ 0 & z_l^{(j)} & 0 \\ i\sqrt{\frac{l+1}{2l+1}}z_{l-1}^{(j)} & 0 & -i\sqrt{\frac{l}{2l+1}}z_{l+1}^{(j)} \end{pmatrix} \begin{pmatrix} \boldsymbol{Y}_{lm}^{l-1} \\ \boldsymbol{Y}_{lm}^{l} \\ \boldsymbol{Y}_{lm}^{l+1} \end{pmatrix} \tag{A1}$$

According to the quantum theory of angular momentum the vector spherical harmonics of Varshalovich et al. [63] can be expressed in spherical basis using Clebsch-Gordan coefficients

$$\boldsymbol{Y}_{jm}^{l}(\hat{\boldsymbol{r}}) = \sum_{m',\sigma} \langle lm'1\sigma|jm\rangle Y_{lm'}(\hat{\boldsymbol{r}})\boldsymbol{e}_\sigma = \sum_{m',\sigma} C_{lm',1\sigma}^{jm} Y_{lm'}\boldsymbol{e}_\sigma$$

where the spherical basis vectors are the eigenstates of spin operators $S^2$ and $S_z$

$$\boldsymbol{e}_{\pm 1} = \mp\frac{1}{\sqrt{2}}(\hat{\boldsymbol{x}} \pm i\hat{\boldsymbol{y}}), \boldsymbol{e}_0 = \hat{\boldsymbol{z}}$$

***Appendix B. Expansion of the Vector Spherical Harmonics over Plane Waves***

The *bianisotropic orbitals* provided by solutions of Eq. (10) can be represented as an expansion over plane wave solutions of the index of refraction operator method [31-33] with the indexes of refraction related with the radial quantum number $k_q$ as $n = k_q/k_0$. This can be understood by inspecting the expansion of the scalar spherical harmonics $\psi_{lm}^{(1)}$

$$\psi_{lm}^{(1)}(k_q;\boldsymbol{r}) = -\frac{1}{i\sqrt{l(l+1)}} j_l(k_q r)Y_{lm}(\hat{\boldsymbol{r}}) = \int \psi_{lm}^{(1)}(k_q;\boldsymbol{k})\, e^{i\boldsymbol{kr}}\frac{d^3k}{(2\pi)^3}$$

According to [63]

$$\psi_{lm}^{(1)}(k_q;\boldsymbol{k}) = \int \psi_{lm}^{(1)}(k_q;\boldsymbol{r})\; e^{-i\boldsymbol{kr}}\,dV = -\frac{(-1)^l}{i\sqrt{l(l+1)}}2\pi^2 i^l\frac{\delta(k_q-k)}{k_q^2}Y_{lm}(\hat{\boldsymbol{k}})$$

which means that all scalar spherical harmonics contain only the plane waves with the wavenumbers $k_q$

$$\psi_{lm}^{(1)}(k_q;\boldsymbol{r}) = -\frac{(-1)^l}{i\sqrt{l(l+1)}}\pi i^l \int e^{ik_q\hat{\boldsymbol{k}}\boldsymbol{r}}\, Y_{lm}(\hat{\boldsymbol{k}})\frac{d^2\Omega_{\boldsymbol{k}}}{(2\pi)^2}$$

Correspondingly, the vector spherical harmonics

$$\boldsymbol{L}_{lm}^{(1)} = \frac{1}{k_q}\nabla\psi_{lm}^{(1)} = -\frac{(-1)^l}{i\sqrt{l(l+1)}}\pi i^l \int e^{ik_q\hat{\boldsymbol{k}}\boldsymbol{r}}\,\hat{\boldsymbol{k}}Y_{lm}(\hat{\boldsymbol{k}})\frac{d^2\Omega_{\boldsymbol{k}}}{(2\pi)^2}$$

$$\boldsymbol{M}_{lm}^{(1)} = \nabla \times \left(\boldsymbol{r}\psi_{lm}^{(1)}\right) = j_l(k_q r)\boldsymbol{X}_{lm}(\hat{\boldsymbol{r}}) = \frac{j_l(k_q r)}{\sqrt{l(l+1)}}\hat{\boldsymbol{L}}Y_{lm} = -\frac{(-1)^l}{i\sqrt{l(l+1)}}\pi i^l \int e^{ik_q\hat{\boldsymbol{k}}\boldsymbol{r}}\,\hat{\boldsymbol{L}}_{\boldsymbol{k}} Y_{lm}(\hat{\boldsymbol{k}})\frac{d^2\Omega_{\boldsymbol{k}}}{(2\pi)^2}$$

$$\boldsymbol{N}_{lm}^{(1)} = \frac{1}{k_q}\nabla \times \boldsymbol{M}_{lm}^{(1)} = \frac{1}{k_q}\nabla \times \left(j_l(k_q r)\boldsymbol{X}_{lm}(\hat{\boldsymbol{r}})\right) = -\frac{(-1)^l}{i\sqrt{l(l+1)}}\pi i^{l-1}\int e^{ik_q\hat{\boldsymbol{k}}\boldsymbol{r}}\,[\hat{\boldsymbol{k}} \times \hat{\boldsymbol{L}}_{\boldsymbol{k}}]Y_{lm}(\hat{\boldsymbol{k}})\frac{d^2\Omega_{\boldsymbol{k}}}{(2\pi)^2}$$

***Appendix C. Expansion of the Vector Plane Waves over Vector Spherical Harmonics***

According to [63]

$$\begin{aligned}\boldsymbol{a}e^{i\boldsymbol{kr}} &= 4\pi\sum_{lm}\sum_{q=l-1}^{l+1} i^q\left\{\boldsymbol{a}\cdot\boldsymbol{Y}_{lm}^{q*}(\hat{\boldsymbol{k}})\right\} j_q(kr)\boldsymbol{Y}_{lm}^{q}(\hat{\boldsymbol{r}}) \\ &= 4\pi\sum_{lm}\Big(i^{l-1}\left\{\boldsymbol{a}\cdot\boldsymbol{Y}_{lm}^{l-1*}(\hat{\boldsymbol{k}})\right\}j_{l-1}(kr)\boldsymbol{Y}_{lm}^{l-1}(\hat{\boldsymbol{r}}) + i^l\left\{\boldsymbol{a}\cdot\boldsymbol{Y}_{lm}^{l*}(\hat{\boldsymbol{k}})\right\}j_l(kr)\boldsymbol{Y}_{lm}^{l}(\hat{\boldsymbol{r}}) \\ &\quad + i^{l+1}\left\{\boldsymbol{a}\cdot\boldsymbol{Y}_{lm}^{l+1*}(\hat{\boldsymbol{k}})\right\}j_{l+1}(kr)\boldsymbol{Y}_{lm}^{l+1}(\hat{\boldsymbol{r}})\Big)\end{aligned}$$

From Eq. (A1)

$$\begin{pmatrix} z_{l-1}^{(j)}\boldsymbol{Y}_{lm}^{l-1} \\ z_{l}^{(j)}\boldsymbol{Y}_{lm}^{l} \\ z_{l+1}^{(j)}\boldsymbol{Y}_{lm}^{l+1}\end{pmatrix} = \begin{pmatrix} -\frac{il\sqrt{l+1}}{\sqrt{2l+1}} & 0 & -\frac{i\sqrt{l+1}}{\sqrt{2l+1}} \\ 0 & 1 & 0 \\ -\frac{i\sqrt{l}(l+1)}{\sqrt{2l+1}} & 0 & i\sqrt{\frac{l}{2l+1}}\end{pmatrix}\begin{pmatrix}\boldsymbol{L}_{lm}^{(j)} \\ \boldsymbol{M}_{lm}^{(j)} \\ \boldsymbol{N}_{lm}^{(j)}\end{pmatrix}$$

$$\begin{aligned}\boldsymbol{a}e^{i\boldsymbol{kr}} = 4\pi\sum_{lm} i^l\Bigg(&\left\{\boldsymbol{a}\cdot\boldsymbol{Y}_{lm}^{l-1*}(\hat{\boldsymbol{k}})\right\}\left(-\frac{l\sqrt{l+1}}{\sqrt{2l+1}}\boldsymbol{L}_{lm}^{(1)} - \frac{\sqrt{l+1}}{\sqrt{2l+1}}\boldsymbol{N}_{lm}^{(1)}\right) + \left\{\boldsymbol{a}\cdot\boldsymbol{Y}_{lm}^{l*}(\hat{\boldsymbol{k}})\right\}\boldsymbol{M}_{lm}^{(1)} \\ &+ \left\{\boldsymbol{a}\cdot\boldsymbol{Y}_{lm}^{l+1*}(\hat{\boldsymbol{k}})\right\}\left(\frac{\sqrt{l}(l+1)}{\sqrt{2l+1}}\boldsymbol{L}_{lm}^{(1)} - \sqrt{\frac{l}{2l+1}}\boldsymbol{N}_{lm}^{(1)}\right)\Bigg)\end{aligned}$$

$$\begin{aligned}= 4\pi\sum_{lm} i^l\Bigg(&-\sqrt{l(l+1)}\boldsymbol{a}\cdot\left(\frac{\sqrt{l}}{\sqrt{2l+1}}\boldsymbol{Y}_{lm}^{l-1*}(\hat{\boldsymbol{k}}) - \frac{\sqrt{l+1}}{\sqrt{2l+1}}\,\boldsymbol{Y}_{lm}^{l+1*}(\hat{\boldsymbol{k}})\right)\boldsymbol{L}_{lm}^{(1)} + \left\{\boldsymbol{a}\cdot\boldsymbol{Y}_{lm}^{l*}(\hat{\boldsymbol{k}})\right\}\boldsymbol{M}_{lm}^{(1)} - \boldsymbol{a} \\ &\cdot\left(\frac{\sqrt{l+1}}{\sqrt{2l+1}}\boldsymbol{Y}_{lm}^{l-1*}(\hat{\boldsymbol{k}}) + \frac{\sqrt{l}}{\sqrt{2l+1}}\boldsymbol{Y}_{lm}^{l+1*}(\hat{\boldsymbol{k}})\right)\boldsymbol{N}_{lm}^{(1)}\Bigg)\end{aligned}$$

$$\boldsymbol{Y}_{lm}^{(-1)}(\hat{\boldsymbol{k}}) = \frac{\sqrt{l}}{\sqrt{2l+1}}\boldsymbol{Y}_{lm}^{l-1}(\hat{\boldsymbol{k}}) - \frac{\sqrt{l+1}}{\sqrt{2l+1}}\,\boldsymbol{Y}_{lm}^{l+1}(\hat{\boldsymbol{k}}), \boldsymbol{Y}_{lm}^{(+1)}(\hat{\boldsymbol{k}}) = \left(\frac{\sqrt{l+1}}{\sqrt{2l+1}}\boldsymbol{Y}_{lm}^{l-1}(\hat{\boldsymbol{k}}) + \frac{\sqrt{l}}{\sqrt{2l+1}}\boldsymbol{Y}_{lm}^{l+1}(\hat{\boldsymbol{k}})\right)$$

We arrive at

$$\boldsymbol{a}e^{i\boldsymbol{kr}} = 4\pi\sum_{lm} i^l\left(-\sqrt{l(l+1)}\left\{\boldsymbol{a}\cdot\boldsymbol{Y}_{lm}^{(-1)*}(\hat{\boldsymbol{k}})\right\}\boldsymbol{L}_{lm}^{(1)} + \left\{\boldsymbol{a}\cdot\boldsymbol{Y}_{lm}^{l*}(\hat{\boldsymbol{k}})\right\}\boldsymbol{M}_{lm}^{(1)} - \left\{\boldsymbol{a}\cdot\boldsymbol{Y}_{lm}^{(+1)*}(\hat{\boldsymbol{k}})\right\}\,\boldsymbol{N}_{lm}^{(1)}\right)$$

## References

[1] K. Schwab, “The fourth industrial revolution,” Currency, 2017.

[2] S. Dang, A. Osama, B. Shihada, M.-S. Alouini. “What should 6G be?” Nature Electronics, 3(1), 20-29 (2020)

[3] E. L. Wolf, M. Medikonda, “Understanding the nanotechnology revolution,” John Wiley & Sons, 2012.

[4] C. Furse, D. A. Christensen, C H. Durney, “Basic introduction to bioelectromagnetics,” CRC press, 2009.

[5] R. A. Poisel, “Information warfare and electronic warfare systems,” Artech House, 2013.

[6] M. I. Jordan, T. M. Mitchell, “Machine learning: Trends, perspectives, and prospects,” Science 349 (6245), 255-260 (2015)

[7] H. Bhaumik, D. Hexner, “Loss of material trainability through an unusual transition,” Physical Review Research 4(4), L042044 (2022)

[8] D. J. Griffiths, “Introduction to Electrodynamics”, 4th ed. (2021).

[9] H. W. Meyer, “A history of electricity and magnetism,” 1971.

[10] E. Whittaker, “A History of the Theories of Aether and Electricity: Vol. I: The Classical Theories; Vol. II: The Modern Theories, 1900-1926,” Dover Publications, 2020.

[11] N. Engheta, R. W. Ziolkowski, eds., “Metamaterials: physics and engineering explorations,” (John Wiley & Sons, 2006)

[12] M. A. Noginov, G. Dewar, M. W. McCall, N. I. Zheludev, “Tutorials in complex photonic media,” (SPIE press, 2009)

[13] S. A. Tretyakov, “A personal view on the origins and developments of the metamaterial concept,” Journal of optics 19(1): 013002 (2016).

[14] E. O. Kamenetskii, “Chirality, Magnetism and Magnetoelectricity,” Berlin: Springer, 2021.

[15] F. Capolino, ed. “Theory and phenomena of metamaterials,” CRC press, 2017.

[16] M. A. Noginov, V. A. Podolskiy, eds., “Tutorials in metamaterials,” (CRC press, 2011)

[17] C. Simovski, S. Tretyakov, “An Introduction to Metamaterials and Nanophotonics,” (Cambridge University Press, 2020)

[18] T. G. Mackay, A. Lakhtakia, “Electromagnetic anisotropy and bianisotropy: a field guide,” (World Scientific, 2010)

[19] D. K. Cheng, J. A. Kong, “Covariant descriptions of bianisotropic media,” Proceedings of the IEEE, 56(3), 248-251 (1968)

[20] W. C. Röntgen, “Ueber die durch Bewegung eines im homogenen electrischen Felde befindlichen Dielectricums hervorgerufene electrodynamische Kraft” Annalen der Physik, 271(10), 264-270 (1888)

[21] H. A. Wilson, “On the electric effect of rotating a dielectric in a magnetic field.” Philosophical Transactions of the Royal Society of London. Series A, Containing Papers of a Mathematical or Physical Character, 204(372-386), 121-137 (1905).

[22] L. D. Landau, E. M. Lifshitz, “Electrodynamics of Continuous Media. Theoretical Physics. Vol. 8”, §51 (Fizmatlit, 2005)

[23] I. E. Dzyaloshinskii, J. Exp. Theoret. Phys. 37, 881 (1959) [translation: Soviet Phys.-JETP 10, 628 (1960)].

[24] I. Lindell, A. Sihvola, S. Tretyakov, A. J. Viitanen, “Electromagnetic waves in chiral and bi-isotropic media,” (Artech House, 1994)

[25] E. O. Kamenetskii, “Bianisotropics and electromagnetics,” arXiv cond-mat/0601467 (2006).

[26] A. Sihvola, I. Semchenko, S. Khakhomov, “View on the history of electromagnetics of metamaterials: Evolution of the congress series of complex media,” Photonics and Nanostructures-Fundamentals and Applications 12, no. 4, 279-283 (2014)

[27] S. A. Tretyakov, F. Bilotti, A. Schuchinsky, "Metamaterials Congress Series: Origins and history," In 2016 10th International Congress on Advanced Electromagnetic Materials in Microwaves and Optics (METAMATERIALS), pp. 361-363. IEEE, 2016.

[28] J. A. Kong, "Theorems of bianisotropic media," Proceedings of the IEEE, 60(9), 1036-1046 (1972).

[29] M. Berry,"The optical singularities of bianisotropic crystals," Proceedings of the Royal Society A: Mathematical, Physical and Engineering Sciences 461(2059), 2071-2098 (2005).

[30] T. Mulkey, J. Dillies, M. Durach, "Inverse problem of quartic photonics," Optics letters, 43(6), 1226-1229 (2018).

[31] M. Durach, R. F. Williamson, M. Laballe, T. Mulkey, "Tri-and tetrahyperbolic isofrequency topologies complete classification of bianisotropic materials," Applied Sciences, 10(3), 763 (2020)

[32] M. Durach, "Tetra-hyperbolic and tri-hyperbolic optical phases in anisotropic metamaterials without magnetoelectric coupling due to hybridization of plasmonic and magnetic Bloch high-k polaritons," Optics Communications, 476, 126349 (2020).

[33] M. Durach, R. Williamson, J. Adams, T. Holtz, P. Bhatt, R. Moreno, F. Smith, "On Fresnel-Airy Equations, Fabry-Perot Resonances and Surface Electromagnetic Waves in Arbitrary Bianisotropic Metamaterials," Progress In Electromagnetics Research, Vol. 173, 53-69 (2022)

[34] M. LaBalle, M. Durach, "Additional waves and additional boundary conditions in local quartic metamaterials," OSA Continuum 2(1), pp. 17-24 (2019)

[35] M. Durach, "Complete 72-parametric classification of surface plasmon polaritons in quartic metamaterials," OSA Continuum 1(1), pp. 162-169 (2018)

[36] D. K. Cheng, J. A. Kong, "Time-Harmonic Fields in Source-Free Bianisotropic Media," Journal of Applied Physics, 39(12), 5792-5796 (1968)

[37] J. A. Kong, "Electromagnetic wave theory," (Wiley-Interscience, 1990)

[38] G. Mie, "Beitraege zur Optik trueber Medien, speziell kolloidaler Metalloesungen," Ann. Phys. Lpz. 25(3), 377–445 (1908)

[39] C.F. Bohren, D.R. Huffman, "Absorption and Scattering of Light by Small Particles," Wiley, 1983

[40] C.F. Bohren, "Light scattering by an optically active sphere," Chem. Phys. Lett. 29(3), 458–462 (1974)

[41] C.-W. Qiu, L.-W. Li, T.-S. Yeo, S. Zouhdi, "Scattering by rotationally symmetric anisotropic spheres: Potential formulation and parametric studies," Phys. Rev. E 75(2), 026609 (2007)

[42] A.D.U. Jafri, A. Lakhtakia, "Scattering of an electromagnetic plane wave by a homogeneous sphere made of an orthorhombic dielectric–magnetic medium," J. Opt. Soc. Am. A 31(1), 89–100 (2014)

[43] Z. Lin, S. T. Chui, "Electromagnetic scattering by optically anisotropic magnetic particle," Physical Review E 69(5) 056614 (2004)

[44] J. L.-W. Li, W.-L. Ong, "A new solution for characterizing electromagnetic scattering by a gyroelectric sphere," IEEE Transactions on Antennas and Propagation 59(9): 3370-3378 (2011).

[45] J. L.-W. Li, W.-L. Ong, K. H. R. Zheng, "Anisotropic scattering effects of a gyrotropic sphere characterized using the T-matrix method," Physical Review E 85(3), 036601 (2012)

[46] A. Novitsky, A.S. Shalin, A.V. Lavrinenko, "Spherically symmetric inhomogeneous bianisotropic media: Wave propagation and light scattering," Phys. Rev. A 95(5), 053818 (2017)

[47] A. Lakhtakia, "New principle for scattering inside a Huygens bianisotropic medium," J. of Optics, 1-9 (2022)

[48] S. S. Kruk, J. W. Zi, E. Pshenay-Severin, K. O'Brien, D. N. Neshev, Y. S. Kivshar, and X. Zhang, "Magnetic hyperbolic optical metamaterials," Nature Commun., Vol. 7, No. 1, 1–7, 2016.

[49] V. R. Tuz, I. V. Fedorin, and V. I. Fesenko, "Bi-hyperbolic isofrequency surface in a magneticsemiconductor superlattice," Optics Letters, Vol. 42, 4561, 2017.

[50] Tuz, V. R. and V. I. Fesenko, “Magnetically induced topological transitions of hyperbolic dispersion in biaxial gyrotropic media,” Journal of Applied Physics, Vol. 128, 013107, 2020.

[51] Z. Guo, H. Jiang, and H. Chen, “Hyperbolic metamaterials: From dispersion manipulation to applications,” Journal of Applied Physics, Vol. 127, No. 7, 071101, 2020.

[52] O. Takayama, A. V. Lavrinenko, “Optics with hyperbolic materials,” JOSA B, Vol. 36, No. 8, F38–F48, 2019.

[53] M. V. Davidovich, “Hyperbolic metamaterials: production, properties, applications, and prospects,” Physics-Uspekhi 62(12), 1173 (2019)

[54] D. R. Smith, D. Schurig, “Electromagnetic wave propagation in media with indefinite permittivity and permeability tensors,” Phys. Rev. Lett. 90, 077405 (2003).

[55] A. Alù, N. Engheta, “Achieving transparency with plasmonic and metamaterial coatings,” Physical Review E 72(1), 016623 (2005)

[56] R. Hodges, C. Dean, M. Durach, “Optical neutrality: invisibility without cloaking,” Optics letters 42(4), 691-694 (2017)

[57] A. Poddubny, I. Iorsh, P. Belov, Y. Kivshar, “Hyperbolic metamaterials,” Nature Photonics 7, 948-957 (2013)

[58] P. Shekhar, J. Atkinson, and Z. Jacob, “Hyperbolic metamaterials: fundamentals and applications,” Nano Convergence 1, 14 (2014)

[59] L. Ferrari, C. Wu, D. Lepage, X. Zhang, Z. Liu, “Hyperbolic metamaterials and their applications,” Progress in Quantum Electronics 40, 1-40 (2015)

[60] M. Poleva, K. Frizyuk, K. Baryshnikova, A. Evlyukhin, M. Petrov, A. Bogdanov, “Multipolar theory of bianisotropic response of meta-atoms,” Physical Review B 107(4), L041304 (2023)

[61] J. A. Stratton, “Electromagnetic theory,” John Wiley & Sons, 2007

[62] J. D. Jackson, “Classical electrodynamics,” 3rd ed., John Wiley & Sons, 1999

[63] D. A. Varshalovich, A. N. Moskalev, V. K. Khersonskii, “Quantum theory of angular momentum,” World scientific, 1988

[64] Y.-L. Geng, X.-B. Wu, L.-W. Li, B.-R. Guan, “Mie scattering by a uniaxial anisotropic sphere,” Physical Review E 70(5), 056609 (2004)

[65] Y.-L. Geng, C.-W. Qiu, “Extended Mie theory for a gyrotropic-coated conducting sphere: An analytical approach,” IEEE Transactions on Antennas and Propagation 59(11), 4364-4368 (2011)

[66] D. Sarkar and N. J. Halas, “General vector basis function solution of Maxwell’s equations,” Phys. Rev. E 56, 1102 (1997).

[67] A. Lakhtakia, “Polarizability dyadics of small bianisotropic spheres,” Journal de Physique 51(20): 2235-2242 (1990)

[68] A. Sihvola, “On polarizability properties of bianisotropic spheres with noncomplete magnetoelectric dyadic,” Microwave and Optical Technology Letters 7(14), 658-661 (1994)

[69] A. Sihvola, “Rayleigh formula for bianisotropic mixtures,” Microwave and Optical Technology Letters 11(2), 73-75, (1996)